\documentclass[9pt]{article}
\usepackage{amsmath,amssymb,amsthm}

\usepackage{graphicx}

\usepackage{amsmath,amsthm,amssymb,caption}

\usepackage{subcaption}
\usepackage{multirow}
\usepackage{tabularx}
\usepackage{arydshln}
\usepackage{comment}
\usepackage{amsmath}
\usepackage{amsthm, amssymb}
\usepackage{graphicx, ctable, url}

\newtheorem{prop}{Proposition}[section]

\newtheorem{exam}[prop]{Example}
\newtheorem{exer}[prop]{$\spadesuit$ Exercise}
\newtheorem{exereind}[prop]{Question}

\theoremstyle{definition}

\newcommand{\be}{\begin{equation}}
\newcommand{\ee}{\end{equation}}
\newcommand{\bes}{\begin{eqnarray*}}
\newcommand{\ees}{\end{eqnarray*}}
\newcommand{\bex}{\begin{exam}\rm}
\newcommand{\eex}{\ \qed\end{exam}}
\newcommand{\boef}{\begin{exer}\rm}
\newcommand{\eoef}{\end{exer}}
\newcommand{\bq}{\begin{exereind}\rm}
\newcommand{\eq}{\end{exereind}}
\newcommand{\bpr}{\begin{proof}}
\newcommand{\epr}{\end{proof}}

\newcommand{\TND}{{\rm test-negative design }}

\begin{document}

\title{Some reflections on the test-negative design}
\author{Ronald Meester and Jan Bonte}
 
\maketitle

\begin{abstract}
We discuss some philosophical, methodological and practical problems concerning the use of the \TND for COVID-19 vaccines. These problems limit the use of this design considerably.
\end{abstract}

\medskip\medskip\noindent
The test-negative design was developed as a method to estimate influenza vaccine effectiveness, but has more recently also been used in the context of COVID-19 vaccines \cite{jackson, fukushima, benn, vandenb, aaby, gazit, sko, che, lopez}. In this contribution, we discuss the vaccine effectivity as defined through a \TND study.  Clearly, in order to do so, we need to specify what ``effectivity'' means, and we discuss this below. The literature on this subject is rather confusing, as we will explain. First we explain the design.
 
In the test-negative design, it is registered how many people with a influenza-like illness request medical care. It may concern people who report to the general practitioner or at a clinic, but it can also concern people who have to be admitted to hospital because of their complaints. After people report to a doctor or health facility, a test is done to determine whether someone has an infection with influenza or not. The group of people who test positive for influenza forms the research group, the group of people who test negative forms the control group. Next, the ratio between vaccinated and unvaccinated in the group of people who test positive for influenza is compared with the ratio between vaccinated and unvaccinated in the group of people who test negative.
  
The difference between a standard case control and the \TND is that in a test-negative design, the study population is prospectively defined by the occurrence of respiratory infection and the subsequent seeking of medical help. In this study design, it is first decided whether or not a person belongs to the study population, based on the complaints and symptoms, and only then is it decided by means of a diagnostic test who belongs to the research group (those with a positive test result), and who belongs to the control group (those with a negative test). In a case control design on the other hand, it is the positive test that determines whether someone belongs to the study group, and those individuals are compared with people who have a negative test, the control group. Thus, in that design the test result is leading, and not the disease symptoms. 

In the studies \cite{sko, che, lopez} mentioned above, the data are collected retrospectively by querying large databases. Retrospective is a characteristic of a case control design, not of a test-negative design. Some studies use both terms, ``test-negative'' and ``case control'', as for instance in \cite{lopez}. This is confusing: it is not a test-negative study, it is a case control study, with all its possible forms of bias of results.

We start the discussion by following \cite{jackson} from which we copy the table below. In the table, the population is divided into appropriate cells:\footnote{In \cite{jackson} writes ``infected'', where we write ``positive'' in the table. Since the decision is made on the basis of a test, our formulation seems more appropriate, and it will play a role later in our contribution.} 

\medskip\medskip\noindent
\begin{center}
\begin{tabular}{|l|c c c c|c c c c|}
\hline
& \multicolumn{4}{c|}{Do seek care for ARI} & \multicolumn{4}{c|}{Do not seek care for ARI} \\
\hline
&  & positive & & & & positive & & \\
& positive  & with & & & positive &  with & &  \\
 & with & other & not & &  with & other & not & \\
 & influenza & pathogen & positive & total & influenza & pathogen  & positive & total\\
\hline
Vaccinated & A & B & C & $N_1$ & D & E & F & $N_2$\\
\hline
Not vaccinated & G & H & I & $N_3$ & J & K & L & $N_4$\\
\hline
\end{tabular}
\end{center}

\medskip\medskip
The quantity of interest, the vaccine effectivity, is now defined as
\begin{equation}
\label{VE}
VE = 1- \frac{(\frac{A}{N_1})}{(\frac{G}{N_3})}.
\end{equation}
Still following \cite{jackson}, this $VE$ could be estimated as (writing lower case letters to indicate samples from the populations described in the table)
\begin{equation}
\label{VEhat}
\hat{VE} = 1-\frac{(\frac{a}{n_1})}{(\frac{g}{n_3})}.
\end{equation}
According to \cite{jackson}, this quantity provides an unbiased estimate of the parameter of interest \eqref{VE}.

Now note that the quantities $A, B, C, G, H$ and $I$ are observed in the \TND study: these count the people that seek help, are tested and subsequently registered according to their test- and vaccination status. However, this is not appreciated in \cite{jackson}. Indeed, they write that ``[...] this design is impossible to implement due to the difficulty of distinguishing persons in cell $C$ from cells $D - F$ and persons in cell $I$ from cells $J - L$ (namely uninfected persons that would have sought care had they developed an ARI).'' Hence, \cite{jackson} seems to confuse ``Do seek care'', with ``Would seek care''. If that were the case, then $N_1$ and $N_3$ would not be observable, and $\hat{VE}$ would {\em not} be an estimator in any statistical sense. They get around this by {\em assuming} that
\begin{equation}
\label{condition}
\frac{B}{N_1}=\frac{H}{N_3},
\end{equation}
in which case $VE$ can be rewritten as
$$
VE = 1- \frac{AH}{GB},
$$
and this can be estimated by
$$
\hat{VE} = 1-\frac{a h}{g b}.
$$
Their assumption is, then, that among persons who would seek care for ARI, the incidence of ARI due to other respiratory pathogens does not differ between vaccinated and unvaccinated persons. However, we noticed that this assumption is not necessary if one wants to estimate $V$ with 
\eqref{VEhat}. The authors in \cite{jackson} apparently assume that those seeking help for respiratory infection but testing negative for both influenza and other pathogens cannot be defined, but they can. Because, when diagnosing for respiratory pathogens, a standard panel is generally used, with a limited number of pathogens. There is no exhaustive testing for all possible pathogens in the clinical situation, simply because that is not cost-effective and has no added value. However, this is not to say that those people who test negative for both influenza and other pathogens are not infected. After all, they do have the clinical picture of respiratory infection? It is quite possible that the pathogen in them was not in the standard panel. All this means that all the people belonging to groups $N_1$ and $N_3$ can simply be counted. And so the assumption made by the authors to arrive at the formula in \eqref{condition} is not necessary at all.

Furthermore, it is worth remembering that to calculate the $VE$ it matters which control group one takes. Here one has three choices: one can take the group that tests positive for a different pathogen, or take the group that tests negative for both influenza and other pathogens, the ``pan-negatives'' or take both groups together as a control. If it turns out that vaccination against a specific pathogen does affect the probability of becoming ill from another pathogen, as suggested in \cite{benn}, then the outcome of the calculation will be affected by the choice of control group. In \cite{fukushima} this is discussed in detail. Next we make some observations concerning this design.

\medskip\noindent
{\em Not useful for overall mortality}\\
We first observe that the \TND is useless when used as a measure for the vaccine effect on overall mortality. Indeed, in order to participate in a research according to the \TND, one must, to begin with, be in principle able to seek medical care. So, if one wants to investigate mortality as an adverse event related to the vaccine, the \TND cannot possibly be useful, and research into the question whether or not a vaccine has an effect on the overall mortality rate should not be carried out with the test-negative design. For instance, elderly and vulnerable people who fall ill and die at home fall out of the scope of the study, while this group might be the one most at risk as far as vaccinations are concerned. 

\medskip\noindent
{\em The idealized character of the design}\\
We next point out the idealized character of the population in the table. It seems highly questionable to assume that such a simple subdivision into sub-populations is realistic. As also discussed in \cite{jackson}, the care-seeking behavior of people can most certainly not be described by a simple dichotomy as discussed above. The question whether or not people seek care may depend, among many other things, on the severeness of the symptoms, on the social pressure to do so, on their vaccination status, on the time of the year, and on many other factors. This observation implies that $VE$ is not a physically existing quantity that we try to approximate as well as we can. It is not a characteristic of the vaccine that we try to discover. Instead, \eqref{VEhat} can only be properly understood as an estimator of the quantity in \eqref{VE} within a very unrealistic and extremely simplistic model of the population. It tells us what the $VE$ {\em would} be in certain circumstances, but not what it {\em is}. This helps explaining why estimates of $VE$ can strongly differ among various studies. Any estimate of $VE$ with the \TND is a joint effect of vaccine properties and circumstances like the actual composition of the sampled population; see also \cite{fukushima}. 

\medskip\noindent
{\em No truth-status}\\
Perhaps it helps to draw an analogy with another statistical concept: the likelihood ratio, see \cite{meester_slooten}. A likelihood ratio tells us which out of two competing hypotheses explains the data best, by computing the probability of the data under each of the hypotheses. This probability is computed in a statistical model with typically many uncertainties about, for instance, parameter values. The likelihood ratio reflects evidential value of the data based on our current knowledge. If we would receive more information about the values of the parameters, then the ensuing likelihood ratio would change. However, there is no such thing as a ``true'' likelihood ratio which would result upon having perfect knowledge of the parameters. The evidential value is not a quantity that resides somewhere and that we need to estimate as best as we can. On the contrary, it is a number that reflects what we can possibly know now, in the given circumstances. What people sometimes call the ``true likelihood ratio'', is a number that we would obtain given perfect knowledge. There is nothing ``true'' about this though, since it does not correspond to any realistic state of affairs. Similarly, the $\hat{VE}$ as obtained in the \TND has no truth value other than that this number says something about what the vaccine may or may not achieve in a situation that we are not in. As in the case of the likelihood ratio, the idealized situation is not interesting if we are far away from it. In this connection, `bias' has to do with how far we are from such an idealized situation: it is an adverb connected to the situation, not to the estimator.   

These remarks mean that indeed we should be very careful with qualifications of the estimator in \eqref{VEhat} as being `(un)biased', a terminology that is used in many discussions \cite{jackson, vandenb}. Indeed, being (un)biased would imply that the estimator has or does not have a systematic deviation from the `true' value. However, we just noticed that such a true value has no meaningful realistic interpretation. 
In \cite{fukushima}, for instance, there is a discussion concerning the ``true $VE$'', but this is in the context of a simulation. In a simulation, one can realize a population as described in the table above, but this is only an indication that such a simulation will have a very limited relation to reality and that we can not draw strong conclusions from it.

\medskip\noindent
{\em The test-negative design for COVID-19 vaccines}\\
Although the \TND was originally developed in the context of influenza, it has more recently been used in the COVID-19 context as well, as a study into the effectiveness of the various vaccines (see earlier references). For such studies, it is important to distinguish between various outcomes that can be the subject of research when it comes to vaccination effectiveness. In \cite{ion}, various outcomes are discussed: infection, hospitalization, and death. We already discussed the problems around overall mortality with this design. In view of the discussions around the COVID-19 vaccines, we add transmission of the virus to this list. Let us comment on the usefulness of the \TND for each of these measures.

What about the effect on hospitalization? If the test-negative design is used to study the effect on hospital and ICU admission, it is essential that the clinical picture fits COVID-19 and that this clinical picture corresponds in terms of complaints and symptoms to the people who test negative for SARS-CoV-2. The indication for admission to the ICU will often be the need for mechanical or artificial ventilation, by which both the research group and the control group are fairly easy to define. But also for hospitalization, the reason for admission must be a clinical picture that fits well with COVID-19 and corresponds well with the clinical picture of people who test negative. As soon as people with a positive test result for SARS-CoV-2 are included with a clinical picture that does not fit COVID-19, the study is clearly disabled. 

A complicating factor is that, certainly in the case of an ICU admission and often also in the case of a hospital admission, the test result is already known before the admission follows. That is a potential source of bias (in the sense described above), because it can influence the description of the clinical picture. And the vast majority of vaccine effectiveness studies are conducted by retrospectively questioning databases, with the description of the clinical picture being one of the selection criteria. 

The research into the effect of vaccination on infection and transmission using the test-negative design deserves a separate discussion, because here a number of factors play a role that are not or much less decisive in the other outcome measures. So far, we haven't said anything about the diagnostic tool used. But whether it is the antigen rapid test or the rtPCR, both tests will perform better in a clinical setting like a hospital than in the general population, because it concerns people with influenza-like symptoms and not asymptomatic people. By ``performing better'' we mean the characteristics of the test, the sensitivity and specificity. What is conveniently ignored in many studies on the effect of vaccination on transmission is that these tests in a largely asymptomatic population are likely to perform much worse than in a selected group people with influenza-like symptoms.  

Another point is that, even if the rtPCR and antigen rapid test would perform the same in an asymptomatic population as in a clinical setting, with a decreasing prevalence (the number of infections), the number of false positive results increases, and can exceed the number of true positive test results. And it goes without saying that the prevalence of infections with SARS-CoV-2 in an asymptomatic population will be (very) much lower than in a group of people with a influenza-like illness. 

To back-up our argument, here are some numerical examples. In the first example, we have 10,000 vaccinated and 10,000 unvaccintaed people. The prevalence of SARS-CoV-19 is 10\% among unvaccinated and 1\% among vaccinated. Tests are perfect. This leads to the following table:

\medskip\medskip
\begin{tabular}{|l|c|c|c|}
\hline
 & test positive & test negative & total \\
 \hline
 vaccinated & $A$ = 100 & $B$ = 9,900 & $N_1$ = 10,000 \\
 \hline
 not vaccinated & $G$ = 1,000 & $H$ = 9,000 & $N_3$ = 10,000 \\
 \hline
 total & 1,100 & 18,900 & 20,000\\
 \hline
\end{tabular}

\medskip\medskip\noindent
This leads to a $\hat{VE}$ of 90\%.

But now assume that the test has a sensitivity of 70\% and a specificity of 95\%. We first compute the number of positive and negative tests among the (un)vaccinated people:

\medskip\medskip
\begin{tabular}{|l|c|c|c|}
\hline
 vaccinated people & infected & not infected & total \\
 \hline
 test positive & 70 & 495 & 565 \\
 \hline
 test negative & 30 & 9,405 & 9,435 \\
 \hline
 total & 100 & 9,900 & 10,000\\
 \hline
\end{tabular}

\medskip\medskip
\begin{tabular}{|l|c|c|c|}
\hline
 unvaccinated people & infected & not infected & total \\
 \hline
 test positive & 700 & 450 & 1,150 \\
 \hline
 test negative & 300 & 8,550 & 8,850 \\
 \hline
 total & 1,000 & 9,000 & 10,000\\
 \hline
\end{tabular}

\medskip\medskip\noindent
These numbers are then substituted in the table to compute the vaccine effectivity:

\medskip\medskip
\begin{tabular}{|l|c|c|c|}
\hline
 & test positive & test negative & total \\
 \hline
 vaccinated & $A$ = 565 & $B$ = 9,435 & $N_1$ = 10,000 \\
 \hline
 not vaccinated & $G$ = 1,150 & $H$ = 8,850 & $N_3$ = 10,000 \\
 \hline
 total & 1,715 & 8,285 & 20,000\\
 \hline
\end{tabular}

\medskip\medskip\noindent
The corresponding $\hat{VE}$ is now reduced to 51\%, as a simple computation shows. If instead we take a sensitivity of 95\% and a specificity of 70\%, then a similar computation shows that $\hat{VE}$ reduces to a mere 16\%. The vaccine effectiveness can also become negative, which is to say that vaccinated people become infected more easily than people who have experienced an infection; see also \cite{ion}. This shows that the calculation of vaccine effectiveness using diagnostic tests with insufficient sensitivity and specificity leads to unacceptable estimates of vaccine effectiveness. In both cases above, it leads to an underestimation of the actual effectiveness. 

It should be clear by now that $\hat{VE}$ in a \TND is problematic for various reasons. The quantity $\hat{VE}$ in the \TND can only properly be interpreted as a quantity in a highly idealized situation which has virtually no relation to reality. The effect of vaccinations, therefore, is difficult to measure or even to define. In any case, $\hat{VE}$ as estimated in the \TND is not a number that can be seen as a very useful characteristic of the vaccine, since it depends on many circumstances. This should be taken into account in every study with the test-negative design.

\end{document}